\documentclass[final]{aipproc}
\usepackage{aipxfm}

\layoutstyle{6x9}
\begin{document}
\bibliographystyle{aipproc}

\title
{\bf MSM Self-Energies at Finite Temperature in the
  Presence of Weak Magnetic Fields: Towards a Full Symmetry Restoration Study}


\author{Mar\'\i a Elena Tejeda-Yeomans}{
  address={Universidad de Sonora, Departamento de F\'\i sica, A. Postal 1626,
  C.P. 83000, Sonora, Mexico},
}

\iftrue

\author{Jorge Navarro}{
  address={Instituto de Ciencias Nucleares, Universidad Nacional Autonoma de Mexico, Apartado Postal 70-543, M\'exico Distrito Federal 04510, Mexico},
  altaddress={Departamento de F\'\i sica, Universidad del Atl\'antico, Km.7 antigua v\'\i a a Puerto Colombia, A. A. 1890, Barranquilla, Colombia},
}

\author{Angel S\'{a}nchez}{
address={Instituto de Ciencias Nucleares, Universidad Nacional Autonoma de Mexico, Apartado Postal 70-543, M\'exico Distrito Federal 04510, Mexico},
altaddress={Instituto de F\'\i sica y Matem\'aticas, Universidad Michoacana de San Nicol\'as Hidalgo, Edificio C-3, Cd. Universitaria C.P. 58040, Michoac\'an, M\'exico},
}

\author{Gabriella Piccinelli}{
  address={Centro Tecnol\'ogico, FES Arag\'on, Universidad Nacional Autonoma de Mexico, Avenida Rancho Seco S/N, Bosques de Arag\'on, Nezahualc\'oyotl, Estado de M\'exico 57130, M\'exico},
}

\author{Alejandro Ayala}{
address={Instituto de Ciencias Nucleares, Universidad Nacional Autonoma de Mexico, Apartado Postal 70-543, M\'exico Distrito Federal 04510, Mexico},
}

\fi

\begin{abstract}
The study of the universe's primordial plasma at high temperature plays an
important role when tackling different questions in cosmology, such as the
origin of the matter-antimatter asymmetry. In the Minimal Standard Model (MSM)
neither the amount of CP violation nor the strength of the phase transition
are enough to produce and preserve baryon number during the Electroweak Phase
Transition (EWPT), which are two of the three ingredients needed to develop
baryon asymmetry. In this talk we present the first part of the analysis done within a scenario
where it is viable to have improvements to the aforementioned situation: we
work with the degrees of freedom in the broken symmetry phase of the MSM and
analyze the development of the EWPT in the presence of a weak magnetic field.
More specifically, we calculate the particle self-energies that include the
effects of the weak magnetic field, needed for the MSM effective potential up to
ring diagrams. 
\end{abstract}

\maketitle
\section{Introduction}
\label{intro}

The dynamics of the Universe's primordial plasma at high 
temperature is at the center of any study performed to elucidate some 
outstanding questions in cosmology such as the origin of matter-antimatter 
asymmetry. Even more, the connection 
of such asymmetry to the occurrence of phase transitions 
in the early evolution of the Universe, has been subject to discussions 
at the core of the mechanisms proposed to generate and sustain this 
matter-antimatter imbalance, such as baryogenesis. The dynamics within each 
mechanism is model dependent, nevertheless as early as
 1967 Sakharov ~\cite{Sakharov:1967dj}
provided three key conditions to develop a baryon asymmetry 
in a universe that started from a symmetric state, namely $(i)$ the existence 
of baryon number\footnote{
$
\frac{\eta_B}{s} = \frac{n_b - n_{\bar b}}{s} \simeq 0.87 \times 10^{-10}
$. See for example ~\cite{Yao:2006px} and references therein.
} violation processes, $(ii)$ C and CP symmetries must also be 
violated and $(iii)$ the system described, should be out of thermal equilibrium.
In spite of the fact that within the Minimal Standard Model (MSM) these 
conditions are met for a first order electroweak phase transition (EWPT), 
the strength of said transition (being weakly of first order $v/T \sim 0.55$ 
for $m_{H}= 0$) 
and the amount of CP violation gained (CP violation in the CKM sector amounts 
to $\eta_B \sim 10^{-20}$) are not enough to produce (nucleosynthesis criteria 
requires $\eta_B \sim 10^{-11}$) and sustain ($v/T \sim 1-1.5$ to avoid 
dilution by the sphaleron process) this asymmetry.

There is a considerable
amount of work providing alternatives that can improve this proposal. In 
particular, taking into account the effects of primordial 
magnetic fields ({\it hypermagnetic} 
fields, before the EWPT) on cosmological processes such as phase transitions, 
has proven to be an effective tool (see for example ~\cite{Kronberg:1993vk,Beck:1995zs,Carilli:2001hj,Barrow:1997mj} and references therein). Among other 
things, 
it is expected that
the EWPT becomes a stronger first order one, just as it happens in the 
superconductive phase transition where an external magnetic field changes 
the order of the phase transition due to the Meissner effect. Understandingly,
it becomes crucial to have the structure of the electroweak effective potential
and to analyze its dependence on the order parameter ($\langle v\rangle$ the Higgs 
vacuum expectation value) on the magnetic field strength, the temperature
and on characteristic energy scales such as the model's masses.

In a recent work ~\cite{Sanchez:2006tt,Ayala:2004dx,Piccinelli:2004eu}
the development of the EWPT in the MSM from the
symmetric phase has been studied in the presence of a weak external
hypermagnetic field up to the contribution of ring diagrams. The main result
is that the presence of the field strengthens the first order
nature of the phase transition. This result is in agreement with
calculations performed at a classical ~\cite{Giovannini:1997eg} and one-loop 
levels ~\cite{Elmfors:1998wz} as well as with lattice 
simulations ~\cite{Kajantie:1998rz}.

In this talk we present the first part of an analysis of the EWPT in
the presence of a constant magnetic field carried out within the degrees of
freedom in the broken symmetry phase where the theory has 
been reduced from $SU(2)_L\times U(1)_Y$ to $U(1)_{\mbox{\small{em}}}$.
More specifically, we calculate the particle self-energies that include the
effects of the weak magnetic field, needed for the MSM effective potential
up to ring diagrams. In our calculations we work in the weak field limit and we keep an arbitrary 
value of the gauge parameter throughout.

Ultimately, we will use these self-energies to calculate the MSM effective 
potential up to the contribution of the \textit{ring} 
diagrams
to study the conditions that would make the EWPT a stronger first order one. 
This together with the details of such calulations, will be presented elsewhere ~\cite{Jorge:2008}.

\section{Charged particle propagating in an external weak magnetic field}

In the broken phase of the Standard Model, there are charged scalars, 
fermions and gauge bosons that couple to the external magnetic field. 
To include the effects of the external field into our description 
of propagating charged particles, we use Schwinger's proper 
time method ~\cite{Schwinger:1951xk}. For brevity we present 
here the generic expression for scalar (${\mathcal P}= D_{B}$) and boson 
propagators (${\mathcal P}=G^{\mu\nu}_{B}$)
(see for example ~\cite{Erdas:1998uu,Chyi:1999fc}):
\begin{eqnarray}
\label{prop}
{\mathcal P}(x,x^{\prime}) &=& \phi(x,x^{\prime}) \int \frac{d^4 k}{(2 \pi)^4} e^{-i
k \cdot (x-x^{\prime})} {\mathcal P}(k)
\end{eqnarray}
where we are considering a vector potential $B_\mu=\frac{B}{2}(0,y,-x,0)$
which generates a constant magnetic field of strength $B$ along the $\hat{z}$ axis.
The gauge dependent phase factor $
\phi (x,x^{\prime}) \equiv \exp\left\{ie \int^x_{x^{\prime}} d\xi^\mu \left[B_\mu 
+\frac{1}{2} F_{\mu\nu} (\xi-x^{\prime})^{\nu}\right]\right\}
$, breaks translation invariance. In fact since the net charge flowing in the loop 
is not zero, this phase does not vanish and should in
principle be included in the computation of self-energies, whenever \eqref{prop}
is used. Nevertheless, we use this expression to compute the
ring contribution to the effective potential which requires closed loop calculations.
In this context, the phase factor becomes the identity and
therefore it does not need to be computed for individual self-energies\footnote{The
same applies for the rest of the self-energy diagrams where there is a net
charge flowing in the loop.}. 

Given that we work with a constant magnetic field along the $\hat{z}$ axis and 
following ~\cite{Ayala:2004dx,Erdas:1998uu}
closely on the notation, the momentum dependent functions $D_{B} (k)$ and $G^{\mu\nu}_{B} (k)$ 
can be written as
\begin{eqnarray}
\label{propSerdas}  
i D_{B} (k)&=&\int_0^\infty \frac{ds}{\cos{eBs}} \exp\left\{ i s
  (k_{||}^2-k_{\bot}^2 \frac{\tan{eBs}}{eBs}-m^2 +i \epsilon)\right\} \\
\label{propBerdas}  
iG^{\mu\nu}_{B}(k) &=&\int_0^\infty \frac{ds}{\cos{eBs}} \exp\left\{ i s
(k_{||}^2-k_{\bot}^2 \frac{\tan{eBs}}{eBs})\right\} \Biggl\{ e^{-is(m^2-i\epsilon)} \left[ -g^{\mu
\nu}_{||}+(e^{2eFs})^{\mu\nu}_{\bot}\right] \Biggr.  \nonumber \\
&& ~~~~~+ \frac{1}{m^2}\left(e^{-is(m^2-i\epsilon)}-e^{-is(\xi m^2-i\epsilon)}\right) \Biggl[\left(k^{\mu}+k_{\lambda}F^{\mu\lambda}((\tan(eBs))/B)\right)
\Biggr.  \nonumber \\
&&~~~~~~\times \left(k^\nu +k_{\rho}F^{\rho\nu}((\tan(eBs))/B)\right) - \Biggl.\Biggl.i\frac{e}{2}\left(F^{\mu\nu}+
g_{\bot}^{\mu\nu}B\tan(eBs)\right)\Biggr]\Biggr\}
\end{eqnarray}
where $(e^{2eFs})^{\mu\nu}_{\bot}= g_{\bot}^{\mu\nu}\cos (2eBs)-F^{\mu\nu}\frac{\sin (eBs)}{B}$ and $m$ represents the corresponding charged scalar and gauge boson mass. Throughout we consider
$g^{\mu\nu}= (1,-1,-1,-1)$ as $g^{\mu\nu}=g^{\mu\nu}_{||}-g_{\bot}^{\mu\nu}$ so that the longitudinal 
and transverse momenta are $k_{||}^2=k_0^2-k_3^2$ and $k_\perp^2=k_1^2+k_2^2$.

It has been shown ~\cite{Chyi:1999fc,D'Olivo:2002sp} that the integral in \eqref{propSerdas} can 
be calculated using Cauchy's theorem with contour deformation, which leads to
\begin{eqnarray}
\label{propSlaguerre}
iD_{B}(k)=2i\sum_{l=0}^{\infty}\frac{(-1)^lL_l(\frac{2k_\perp^2}{e B})
{\mathrm e}^{-\frac{k^2_\perp}{e B}}}{k^2_{||}-(2l+1)e B-m^2+i\epsilon},
\end{eqnarray}
where $L_l$, $L_l^m$ are Laguerre and Associated Laguerre polynomials,
respectively. Performing a similar analysis on \eqref{propBerdas} we obtain the following {\bf new} result ~\cite{Jorge:2008}:
\begin{eqnarray} 
\label{propBlaguerre}
iG_{B}^{\mu\nu}(k)&=& 2i \sum_{\lambda =-1}^{1}\sum_{l=0}^{\infty}
\frac{(-1)^l {\mathrm e}^{-\frac{k^2_\perp}{e B}}}{k^2_{||}-(2l+2\lambda+1)e B-m^2+i\epsilon} \\
&& \times \Biggl[ T^{\mu\nu } +\frac{
  m^2(1-\xi)P^{\mu\nu}}{k_{\parallel}^{2}-(2l+2\lambda+1) eB-\xi
  m^{2}+i\epsilon} \Biggr] L_l\Biggr(\frac{2k_\perp^2}{e B}\Biggl) 
\end{eqnarray}
where 
\begin{eqnarray*}
T^{\mu\nu }&=&\sum_{\lambda = -1}^{1}\Biggl[g^{\mu \nu} \left(\left|
\lambda\right| -1\right) -2 g_{\perp}^{\mu\nu}\frac{\left(\left| 
3\lambda \right| -2\right)} {\left(2i\right)^{2}}+2\frac{F^{\mu \nu }}{B} 
\left( \frac{\lambda }{4i} \right)\Biggr] \\
P^{\mu \nu}&=&\frac{1}{m^2}\Biggl[\Biggl(k^{\mu}k^{\nu}-i\frac{e}{2}F^{\mu \nu} \Biggr)
+\Biggl(\frac{k_\lambda k^\nu F^{\mu\lambda} }{B}+\frac{k_{\rho }k^{\mu }F^{\rho\nu}}{B} 
+i\frac{eB}{2}g_{\perp }^{\mu \nu}\Biggr)(ieB)\frac{d}{dk^2_\perp}\Biggr.  \\
&& ~~~~~~~~~~ + \Biggl. \Biggl(\frac{k_{\lambda }k_{\rho }F^{\mu \lambda } F^{\rho\nu
    }}{B^{2}}\Biggr)(ieB)^2\frac{d^2}{d(k^2_\perp)^2}\Biggr]
\end{eqnarray*}

As was discussed earlier, in this work we want to explore the effects of weak primordial magnetic 
fields on the EWPT. In order to determine the appropriate energy scales during this epoch, we restort
to bounds provided by the analysis of cosmological processes in the early universe.
The simple bound $B<T^2$ is obtained, if we require that the magnetic energy density 
$\rho_{mag} \sim B^2$ should be smaller than the overall radiation energy density 
$\rho_{rad} \sim T^4$ given by nucleosynthesis analysis. Furthermore, the field strength 
is also weak compared to $m_W^2$, if we enforce stability against the formation of 
a $W$-condensate. Therefore, we work explicitly with the assumption that the hierarchy of 
scales $eB \ll m^2 \ll T^2$ is obeyed, were we consider $m$ as a 
generic mass of the problem at the electroweak scale (see for example 
~\cite{Maartens:2000ka,Ambjorn:1988tm}). In this context, to be able
to perform the summation over Landau levels, we can perform a 
weak field expansion and write the propagators as power series in $eB$:
\begin{eqnarray}
\label{propSbdebil}
D_{B}(k) &=& \frac{1}{k^2-m^2}\Biggl[ 1-\frac{ (eB)^2}{(k^2-m^2)^2}-
  \frac{2{ (eB)^2} k_\perp^2}{(k^2-m^2)^3}\Biggr],
\end{eqnarray}
for the charged scalar ~\cite{Ayala:2004dx,Sanchez:2006tt}.

Similarly, for the charged gauge boson we obtain the following {\bf new} 
result ~\cite{Jorge:2008}:
\begin{eqnarray}
\label{propBbdebil}
G_{B}^{\mu\nu}\hspace{-0.3cm}&=& 
-i \left(\frac{g^{\mu\nu}}{k^2-m^2}-(1-\xi)\frac{k^{\mu}k^{\nu}}
   {(k^2-m^2)(k^2-\xi m^2)}\right) \nonumber \\
&& -{\left( eB\right)}\Biggl[
\frac{k_{\rho}}{m^{2}
   }\left(  k^{\nu}\frac{F^{\mu\rho}}{B}+k^{\mu}\frac{F^{\rho\nu}}%
   {B}\right)\left(\frac{1}{\left(k^{2}-m^{2}\right)^{2}}-
   \frac{1}{\left( 
   k^{2}-\xi m^{2}\right)^{2}}\right)\Biggr. \nonumber \\
&& - \Biggr. \frac{F^{\mu\nu}}{B}\left(\frac
   {2}{\left(k^{2}-m^{2}\right)^{2}}+\frac{\left(1-\xi\right)}{2\left(
   k^{2}-m^{2}\right)  \left(k^{2}-\xi m^{2}\right)}\right)\Biggr] \nonumber \\ 
&&+i{ \left(eB\right)^{2}} \Biggl[\frac{g^{\mu\nu}+4g_{\bot}^{\mu\nu}}{\left(
   k^{2}-m^{2}\right)  ^{3}}+\frac{2g^{\mu\nu}k_{\bot}^{2}}{\left(
   k^{2}-m^{2}\right)  ^{4}} \Biggr. 
-\frac{k^{\mu}k^{\nu}}{m^{2}}\left(\frac{1}{\left(
   k^{2}-m^{2}\right)^{3}} -\frac{1}{\left(  k^{2}-\xi
   m^{2}\right)^{3}}\right) \nonumber \\
&&-2\frac{k^{\mu}k^{\nu}}{m^{2}}k_{\bot}^{2}\left(  \frac{1}{\left(
   k^{2}-m^{2}\right)  ^{4}}-\frac{1}{\left(  k^{2}-\xi m^{2}\right)
   ^{4}}\right)  
+\frac{g_{\perp}^{\mu\nu}}{2m^{2}}\left(  \frac{1}{\left(
   k^{2}-m^{2}\right)  ^{2}}-\frac{1}{\left(  k^{2}-\xi
   m^{2}\right)^{2}}\right) \nonumber \\
&&-\left.\frac{2}{m^{2}}\left(
   \frac{k_{\lambda}F^{\mu\lambda}k_{\rho}
   F^{\rho\nu}}{B^{2}}\right)  \left(  \frac{1}{\left(  k^{2}-m^{2}\right)^{3}
   }-\frac{1}{\left(k^{2}-\xi m^{2}\right)^{3}}\right)\right]
\end{eqnarray}

\section{MSM charged particle self-energies in the presence of an 
external weak magnetic field}

In the process of calculating the MSM charged particle self-energies in the 
presence of an external magnetic field, we work in the imaginary time formalism 
of thermal quantum field theory. This amounts to having discrete values for 
boson energies ($k_4=\omega_n=2n \pi T$ with $n$ an integer), which in turn 
affects the way we perform the integration over loop momenta, i.e.
$$
\int \frac{d^4k}{(2\pi)^4} \rightarrow i \int \frac{d^4k_E}{(2\pi)^4}
\rightarrow T \sum_n \int \frac{d^3k}{(2\pi)^3}.
$$
Note that the integration is carried out in Euclidean space with $k_0=ik_4$.
Also, we work within the hard thermal loop approximation (HTL), that is we 
take the loop momentum to be of the same order of $T$ which will be the dominant 
scale ($k \sim T \gg p, m$). Furthermore, the sum over Matsubara frequencies 
and integration over the loop 3-momenta is performed using Bedingham's 
method ~\cite{Bedingham:2000ct}. Finally, it is convenient for our purposes
to work in the {\it infrared limit} ($p_0=0$, ${\mathbf p}\to 0$), which
accounts for the plasma screening properties ~\cite{Weldon:1992bv,Mishaps:2008}.

\begin{figure}
\resizebox{0.6\textwidth}{!}
           {\includegraphics{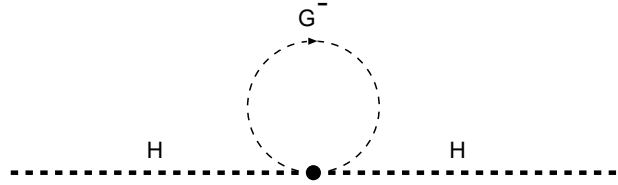}}
 \caption{One of the tadpole diagrams for the Higgs self-energy}
 \label{fig:tadH}
 \source{generated with DIANA ~\cite{Tentyukov:1999is}}
\end{figure}

Based on the MSM Lagrangian (see for example ~\cite{Aoki:1982ed}), we 
generate the diagrams contributing 
to the one-loop self-energy for the photon, Higgs, $W^{\pm}$ and $Z$ 
bosons\footnote{In principle the 
fermion self-energies are also affected by the magnetic field but, 
as in the case of zero external field, their contribution to 
the ring diagrams is subdominant in the infrared (IR) and do not 
need to be taken into account.}.
This means that there are several diagrams to be calculated but due to
lack of space we will present the details elsewhere~\cite{Jorge:2008}. 
Even so, for completeness, we will show the calculation of the
Higgs tadpole diagram shown in fig.(\ref{fig:tadH}), with a charged boson loop. 
This is one of the simplest diagrams to consider with particles 
propagating in the loop which are affected by the external magnetic field:
\begin{eqnarray*}
\Pi_{HG}^{\mbox{tad}} &=&-2i \lambda \int{\mathcal D}{\mathbf k}~
 D_{B}(\omega_n,\mathbf{k};\tilde m)
= 2 \lambda \int{\mathcal D}{\mathbf k}~
\left[\frac{1}{(\omega_n^2+\mathbf{k}^2+ \tilde m^{2})} \right.\\
&&\left. - { (eB)^2} \Biggl(
\frac{1}{(\omega_n^2+\mathbf{k}^2+ \tilde m^{2})^3}
- \frac{2\mathbf{k}_\perp^{2}}{(\omega_n^2+\mathbf{k}^2+ \tilde m^{2})^4}
\Biggr) \right] \\
&=& 2 \lambda \int{\mathcal D}{\mathbf k}~
\Big[I_{1}(\mathbf{k};\tilde{m}^2)-{ (eB)^2}\Big(I_{3}(\mathbf{k};\tilde{m}^2)
- I_{4}(\mathbf{k};\tilde{m}^2)\Big)\Big] \\
&=&\lambda \frac{T^2}{2}  \left(\frac{1}{3}- \frac{\tilde{m}}{%
\pi T} - \frac{(e B)^2}{24 \pi T \tilde{m}^3} \right)
\end{eqnarray*}
where $\tilde{m}$ stands for the Goldstone boson mass and we have made use 
of \eqref{propSbdebil} and performed the
summation over Matsubara frequencies together with integration over 
loop 3-momenta 
($\int {\mathcal D}{\mathbf k} \equiv T \sum_n \int 
\frac{d^3 \mathbf{k}}{(2\pi)^3}$)
using the techniques already mentioned. In all our results for the 
self-energy diagrams, we have kept terms 
representing the leading contributions, namely: terms of
order $(eB)^2/T^4$ can be safely neglected due to the 
aforementioned hierarchy of energy scales and terms of order 
$\tilde{m}/T$ can be safely neglected within HTL.
However, since we are interested in keeping the leading
contribution in the magnetic field strength, we are forced to 
keep terms like $\tilde{m}/T$ which, for a large top 
quark mass (a large $f$) are of the same order as terms 
$(eB)^2/T \tilde{m}^3$. Putting all the diagrams together, 
we get the leading temperature contribution
to the Higgs self-energy at one-loop in agreement with the literature and
the first non-vanishing correction due to the effect of the 
external magnetic field, i.e.
\begin{eqnarray*}
\Pi_{H}&=&\frac{T^2}{4}\Biggl[\left(\frac{3}{4}g^2+\frac{1}{4}g'^2+2\lambda+f^2
\right) + {\mathcal O}\left(\frac{m}{T}\right) + {\mathcal O}\left(\frac{(eB)^2}{Tm^3}\right) \Biggr]
\end{eqnarray*}

Finally, it is worth mentioning that for the $W^{\pm}$ self-energy, the 
calculations are more complicated, specially because there is now a 
richer tensorial structure involved since we
have three independent vectors to our disposal namely $u^\mu$, $q^\mu$ 
and $b^\mu$. Therefore, these self-energies can be written as a linear
combination of nine independent structures ~\cite{D'Olivo:2002sp}. In our case,
only $u^\mu$ and $b^\mu$ remain since we are considering the IR limit.
So, in this limit the structure of the self-energy, collapses to 
$\Pi^{\mu \nu}_{ab}= \Pi^Q_{ab} Q^{\mu\nu} + \Pi^R_{ab} R^{\mu\nu} +
\Pi^S_{ab} S^{\mu\nu} + \Pi^M_{ab} g^{\mu\nu}$
where $Q^{\mu\nu}=u^\mu u^\nu, S^{\mu\nu}=u^\mu b^\nu + u^\nu b^\mu, 
R^{\mu\nu}=b^\mu b^\nu $ and the transversality condition 
$q_\mu\Pi_{ab}^{\mu\nu}=0$ is trivially satisfied. For further
details on this result, please refer to ~\cite{Jorge:2008, Jorge:thesis}.
We perform several checks on our calculations and it is worth emphasizing
that, in the limit of zero external magnetic field, we recover 
results already reported in literature ~\cite{Carrington:1991hz}.
 Moreover it is well known that in the absence of an external magnetic 
fields, the MSM thermal self-energies are gauge independent when 
considering only the leading contributions in 
temperature~\cite{LeBellac:1996}. However, when considering the effects 
of a weak external magnetic field, these self-energies turn out 
to be gauge dependent.

\section{Final remarks}

During this talk we work with the degrees of freedom in the broken 
symmetry phase of the MSM to analyze the development of the EWPT 
in the presence of a weak magnetic field. We calculate the 
particle self-energies that include the effects of the external field.
Still in progress is the ${\mathcal O}[(eB)^2]$ contribution to the 
effective potential, adding the ring corrections and the numerical analysis 
of its behavior near the phase transition. Finally, a study on 
gauge dependence of the effective potential is crucial for further 
developments.

\vskip .5cm
\noindent{\bf Acknowledgments}: The research discussed in this talk
is partially supported by Intercambio UNAM-USON under grant PZ-444, 
DGAPA under grant PAPIIT IN116008, CONACyT under grants 52547-F 
and 40025-F and Universidad del Atl\'antico.

\bibliography{dpyc}

\begin{thebibliography}{27}
\expandafter\ifx\csname natexlab\endcsname\relax\def\natexlab#1{#1}\fi
\providecommand{\enquote}[1]{``#1''}
\expandafter\ifx\csname url\endcsname\relax
  \def\url#1{\texttt{#1}}\fi
\expandafter\ifx\csname urlprefix\endcsname\relax\def\urlprefix{URL }\fi
\providecommand{\eprint}[2][]{\url{#2}}

\bibitem[Sakharov(1967)]{Sakharov:1967dj}
A.~D. Sakharov, \emph{Pisma Zh. Eksp. Teor. Fiz.} \textbf{5}, 32--35 (1967).

\bibitem[Yao et~al.(2006)]{Yao:2006px}
W.~M. Yao, et~al., \emph{J. Phys.} \textbf{G33}, 1--1232 (2006).

\bibitem[Kronberg(1994)]{Kronberg:1993vk}
P.~P. Kronberg, \emph{Rept. Prog. Phys.} \textbf{57}, 325--382 (1994).

\bibitem[Beck and et~al(1996)]{Beck:1995zs}
R.~Beck, and et~al, \emph{Ann. Rev. Astron. Astrophys.} \textbf{34}, 155--206
  (1996).

\bibitem[Carilli and Taylor(2002)]{Carilli:2001hj}
C.~L. Carilli, and G.~B. Taylor, \emph{Ann. Rev. Astron. Astrophys.}
  \textbf{40}, 319--348 (2002).

\bibitem[Barrow et~al.(1997)]{Barrow:1997mj}
J.~D. Barrow, P.~G. Ferreira, and J.~Silk, \emph{Phys. Rev. Lett.} \textbf{78},
  3610--3613 (1997).

\bibitem[Sanchez et~al.(2007)]{Sanchez:2006tt}
A.~Sanchez, A.~Ayala, and G.~Piccinelli, \emph{Phys. Rev.} \textbf{D75}, 043004
  (2007).

\bibitem[Ayala et~al.(2005)]{Ayala:2004dx}
A.~Ayala, A.~Sanchez, G.~Piccinelli, and S.~Sahu, \emph{Phys. Rev.}
  \textbf{D71}, 023004 (2005).

\bibitem[Piccinelli and Ayala(2004)]{Piccinelli:2004eu}
G.~Piccinelli, and A.~Ayala, \emph{Lect. Notes Phys.} \textbf{646}, 293--308
  (2004).

\bibitem[Giovannini and Shaposhnikov(1998)]{Giovannini:1997eg}
M.~Giovannini, and M.~E. Shaposhnikov, \emph{Phys. Rev.} \textbf{D57},
  2186--2206 (1998).

\bibitem[Elmfors et~al.(1998)]{Elmfors:1998wz}
P.~Elmfors, K.~Enqvist, and K.~Kainulainen, \emph{Phys. Lett.} \textbf{B440},
  269--274 (1998).

\bibitem[Kajantie and et~al(1999)]{Kajantie:1998rz}
K.~Kajantie, and et~al, \emph{Nucl. Phys.} \textbf{B544}, 357--373 (1999).

\bibitem[Navarro et~al.(2008)]{Jorge:2008}
J.~Navarro, M.~E. Tejeda-Yeomans, A.~Sanchez, G.~Piccinelli, and A.~Ayala
  (2008), work in progress.

\bibitem[Schwinger(1951)]{Schwinger:1951xk}
J.~S. Schwinger, \emph{Phys. Rev.} \textbf{82}, 914--927 (1951).

\bibitem[Erdas et~al.(1998)]{Erdas:1998uu}
A.~Erdas, C.~W. Kim, and T.~H. Lee, \emph{Phys. Rev.} \textbf{D58}, 085016
  (1998).

\bibitem[Chyi et~al.(2000)]{Chyi:1999fc}
T.-K. Chyi, et~al., \emph{Phys. Rev.} \textbf{D62}, 105014 (2000).

\bibitem[D'Olivo et~al.(2003)]{D'Olivo:2002sp}
J.~C. D'Olivo, J.~F. Nieves, and S.~Sahu, \emph{Phys. Rev.} \textbf{D67},
  025018 (2003).

\bibitem[Maartens(2000)]{Maartens:2000ka}
R.~Maartens, \emph{Pramana} \textbf{55}, 575--583 (2000).

\bibitem[Ambjorn and Olesen(1989)]{Ambjorn:1988tm}
J.~Ambjorn, and P.~Olesen, \emph{Nucl. Phys.} \textbf{B315}, 606 (1989).

\bibitem[Bedingham(2000)]{Bedingham:2000ct}
D.~J. Bedingham  (2000), \eprint{hep-ph/0011012}.

\bibitem[Weldon(1993)]{Weldon:1992bv}
H.~A. Weldon, \emph{Phys. Rev.} \textbf{D47}, 594--600 (1993).

\bibitem[Ayala et~al.(2008)]{Mishaps:2008}
A.~Ayala, G.~Piccinelli, A.~Sanchez, and M.~Tejeda-Yeomans  (2008), work in
  progress.

\bibitem[Aoki and et~al(1982)]{Aoki:1982ed}
K.~I. Aoki, and et~al, \emph{Prog. Theor. Phys. Suppl.} \textbf{73}, 1--225
  (1982).

\bibitem[Tentyukov and Fleischer(2000)]{Tentyukov:1999is}
M.~Tentyukov, and J.~Fleischer, \emph{Comput. Phys. Commun.} \textbf{132},
  124--141 (2000).

\bibitem[Navarro(2008)]{Jorge:thesis}
J.~Navarro  (2008), ph.D. thesis, in progress.

\bibitem[Carrington(1992)]{Carrington:1991hz}
M.~E. Carrington, \emph{Phys. Rev.} \textbf{D45}, 2933--2944 (1992).

\bibitem[Le~Bellac(1996)]{LeBellac:1996}
M.~Le~Bellac  (1996), {Cambridge University Press}.

\end{thebibliography}

\end{document}